\documentclass[11pt]{article}
\usepackage{epsf}
\usepackage{graphicx}        % standard LaTeX graphics tool

\def\ll{\label}

\def\c{\cite}

\def\r1{(\ref{$1})}

\def\ba{\begin{array}{c}}

\def\ea{\end{array}}

\def\l{\left}
\def\l({\left(}
\def\r){\right)}
\def\r{\right}

\def\la{\lambda}

 \def\be{\begin{equation}}
\def\bc{\begin{center}}
\def\ec{\end{center}}
\def\bit{\begin{itemize}}
\def\eit{\end{itemize}}
\def\ee{\end{equation}}
\def\ed{\end{document}}
\def\bea{\begin{eqnarray}}
\def\eea{\end{eqnarray}}
\def\efr{\end{flushright}}

\begin{document}
\title{{ Nonlinearizing linear equations to integrable systems
including new hierarchies with nonholonomic deformations
% with  source terms
}}
%\title{{ Generating   integrable systems by nonlinearizing linear equations }}
\author{ Anjan Kundu\\
 Theory Group \& CAMCS, Saha Institute of Nuclear Physics\\
 Calcutta, INDIA\\
{anjan.kundu@saha.ac.in}} \maketitle
%\newpage
\noindent Running Title: {\it Nonlinearizing linear equations }


%nonlin nonloc FT

\begin{abstract}
 We propose a scheme for nonlinearizing linear equations to
generate integrable nonlinear  systems of both the AKNS and the KN classes, based on
the simple idea of dimensional analysis and detecting the building blocks of
 the Lax pair. Along with the well known equations we discover a novel
integrable hierarchy of  higher order nonholonomic deformations for the
AKNS family, e.g. for the KdV, the mKdV, the NLS and the SG equation, showing
thus a two-fold universality of the recently found
  deformation for the KdV equation \c{6kdv}.
% source equations, with the source    transformable  recursively
%  to the  next higher   order. 
% 47
\end{abstract}

\noindent {\bf I. INTRODUCTION}
%\section{Introduction}
%\label{sec1}

In spite of the   significant achievements in the theory of 
integrable systems  over the last fifty
years, 
% with a    large number   of studies   dedicated  to this subject.
  it   remains a mystery: {\it what is  integrable nonlinearity}, i.e.   
 how to identify a priory
 those nonlinear terms  which
  make an equation  integrable. 

Nevertheless    one  notices that,
although integrable equations can be of diverse types,
the integrability filters out  the nonlinearity  to some
specific forms, as evident from the well known equations
like the nonlinear Schr\"odinger  (NLS) equation,  the derivative  NLS
(DNLS), the
 Korteweg-de Vries (KdV) equation, the modified KdV (mKdV), the sine-Gordon (SG) equation
 etc.\cite{solit1}.
Moreover, 
  all  such integrable nonlinear PDEs  
 exhibit exact  soliton solutions, 
    localized and stable form  of which is believed to be 
due to   a fine balance between their linear dispersive  
 and  nonlinear terms.
Therefore it may not  be unexpected  to  expect that a linear PDE,
the dispersive part of which would balance the   nonlinear term, should 
have the  
information of the  integrable nonlinearity  hidden in it.
Consequently, there must be a  
scheme for nonlinearizing  linear  equations 
for generating   integrable  systems.
However
 attacking 
   this   anti-intuitive question directly 
seems  to have been avoided in the literature,
although  the  Ablowitz-Kaup-Newell-Segur (AKNS) scheme \cite{solit1}
and  the celebrated Sato construction \cite{sato} aimed around 
this line and  a recent attempt
 came close to it
\cite{dimitri07}.

We  propose here a direct  but  easy   
scheme   for  nonlinearizing  the 
     linear equations
 \be iq_t=q_{xx}, \ q_t=q_{xxx}, \ \theta_{xt}=0 , \ll{lineq}\ee
 etc. together with their conjugates, which are 
 in fact the linearized parts of the 
well known  equations like the NLS, the DNLS, the KdV, the SG etc.
Our aim is  to generate    
    first  the known integrable systems,
proving thus the effectiveness of our alternative scheme based on a 
 simple  physical idea of dimensional analysis and then,  continuing
the process    search for
   new integrable  nonholonomic deformations for all known equations.

It is  evident  that  
  the direct   nonlinear extension of  a linear equation
 is not  unique, since it alone can not identify the
integrable
 nonlinearity  and   one   
 needs  some extra structure 
 for filtering out the  integrable cases.
 Our strategy therefore is
 to     stick from the    beginning to a   Lax pair,  
a property    inherent
  to all integrable systems,  by  
  defining a pair of  naive  Lax operators $U^{(l)}, V^{(l)} $
   for  the linear system itself
  and then nonlinearize them  to a genuine 
 pair $ (U , V ) $,
%  satisfying  the Lax equations  $ \Phi_x=U\Phi, \ \Phi_t=V\Phi$, from
 which finally  generate
    the
 nonlinear  integrable equation   through 
%the compatibility $ \Phi_{xt}=\Phi_{tx}$ or
 their  flatness condition
\be U_t- V_x+[U,V]=0 \ll{flatness}\ee.

 It is true that   several sophisticated formal methods like the prolongation
technique, the Painlev\'e truncation  \cite{lakshB}, the Sato theory
\cite{sato}, the AKNS method \cite{solit1}
    etc. are available in the literature 
for constructing the  Lax pair of a nonlinear integrable equation.
 However our alternative construction,  
  based on   a fundamental concept  of
dimensional analysis, is perhaps   the simplest and the most  physical one. 
%Our strategy is to  build up  a corresponding naive {\it linear Lax
%pair} and  deform them nonlinearly to a genuine
%   pair, by using  a simple scaling   dimensional   argument
% together with  a few
%   input.
 Exploiting only the scaling  dimensions
and  identifying  the constituent
 building blocks  of the  Lax operators, hidden already in the linear
equations, 
we can construct  a Lax pair  quite  easily     to 
  any higher order for  both the AKNS
\cite{solit1} and the  Kaup-Newell (KN)  \cite{kn} spectral problems.
Interestingly, by regulating the     scaling dimension of the  field 
 we can 
generate  uniquely both the    NLS
 and the DNLS equations, starting 
from the same  linear Schr\"odinger (LS)
equation.
Finally as an important application  
 we discover  new integrable
hierarchies of nonholonomic    deformations for  the  
well known equations like the KdV, the mKdV, the NLS, the SG etc. The nonholonomic
deformation in this field theoretical context is given by some differential
constraints on the deforming function. A
 recently discovered 6-th order  integrable
KdV (6KdV) equation, attracting  much attention \cite{6kdv,kup08,dkdv08},
happens to be
the  lowest order deformation of this kind
for the KdV equation.
 We  are able to   generalize here  not only the earlier 
result on the deformed KdV and  the 6KdV  by unveiling 
 a novel hierarchy of their higher order deformations, but also to show   the
universality of such perturbations,  by   discovering
  integrable nonholonomic  deformations 
 for all  members  of the AKNS family, 
 including their higher order   integrable hierarchies, a task 
termed  as highly desirable but quite impossible in a
 recent study  \cite{kup08}. 
Interestingly,  such  deformed integrable  systems can  be 
represented also as known  nonlinear equations     with additional perturbative
 terms subjected to
 differential constraints.  Such perturbations, 
contrary to the usual belief, not only preserve the integrability 
of the original equation but also yield richer properties.   
%Nonlinear equations with source terms appear in many physical models
%.

  Nonlinear integrable systems like the
 NLS, the mKdV and the SG equations 
with  various perturbations  
%variety of forcing, damping, inhomogeneities
  appear
in many physical situations, e.g. fluxons in Josephson junction,
    parametrically driven  damped molecular chain, 
nonlinear optical fiber  communication, ferromagnet in variable magnetic
field, nonlinear Faraday resonance etc. \cite{sourceNI}.
 However in almost all these perturbed systems the integrability - the most
cherishable property of a nonlinear system - 
 is usually lost.  As a result all their related precious properties
 like higher conserved quantities,  exact soliton
solutions, elastic scattering of solitons etc., 
 enormously useful in physical applications, are also
lost. Therefore constructing  integrable perturbed systems, preserving their
complete integrability and exact soliton solutions, is a challenging physical
problem, which we are able to solve here  to certain extend.

Our integrable perturbed equations 
exhibit the usual  exact N-soliton
solutions, but with   an unusual  accelerated or decelerated  motion.
 Recall
that the standard  solitons   move   with a  constant velocity  and
  behave like { particles} undergoing elastic
collisions with other solitons, which  have been of significant importance and
application in various physical problems. In the present context of integrable perturbed
equations the accelerated  solitons behave like particles under   external
 forces. They     are subjected
to perturbations, which are themselves   solitonic  in nature. In fact
  the time-dependent 
 asymptotic value of such a perturbing  function
% $c(t)$
acts like a force  sitting at the space boundaries and controlling  the
  motion of the field  soliton, depending
on the nature of which
%(i.e. $ c>0$ or $c<0 $)
 the soliton can  accelerate
or decelerate.

 Accelerated solitons  appear  in many physical models like inhomogeneous
plasma \cite{PRL76} 
%[PRL,37,693 (1976)]
, information transfer in the DNA chain \cite{PRA91}
% [PR A 44, 5292 (1991)]
,  transport of solitons and bubbles \cite{CSF06}
% [Chaos Solitons & Fractals, 28,804(2006)]
 etc. Therefore the present integrable  models
 with  exact  accelerating  solitons  showing
elastic soliton scattering \cite{kunjpa082} could find applications in practical
 situations.

%Before proceeding  to our   actual  construction using SD argument,
%let us see how it works
% in   estimating     nonlinear terms addable  to a linear equation like $
%iq_t=q_{xx}$, where due to the length dimension or the  SD of
%x-derivative being  $[\frac \partial {\partial x}]=[  L]^{-1}$,
% the  SD of each
% term  is $[q][  L]^{-2}$. Assuming    SD of the field to be  $[q]=[  L]^{-1} $ and
%  the nonlinear coupling dimension-less, the allowed
%   additional nonlinear terms with the same SD are
%$|q|^2q, |q|^2 q^*, qq^*_x, q^*q_x, q^3,(q^*)^3$, of which only the
%first possibility is consistent with the integrable  NLS equation.
%Similar situation but with  more  ambiguity arises  in higher order
%linear  equations, concluding
% Noticing that   direct   nonlinearization of  a linear equation
% is not  unique and one needs a more refined
%method for filtering out the integrable cases,
%  we  stick from the    beginning to a  structure akin
%  to  integrable systems   by  defining
%a  Lax pair for the linear system itself  and proceed to
%nonlinearize them  to  a true    pair $ (U , V ) $

The arrangement of  the paper is  as follows. Sec. II
 introduces our nonlinearization
scheme.
% with  the idea of scaling dimension and building blocks of the Lax pair explained in Sec. III.
  Generation of specific  nonlinear integrable systems is
  presented in Sec. III.
 New
integrable nonholonomic deformations  of the AKNS system are detailed in Sec. 
IV,
   with the concrete
examples given in Sec. V. In Sec. VI the present result
is listed  and compared with the 
  existing ones.  Sec. VII is  the concluding section
 followed by the
bibliography.
\vskip .8cm

\noindent{\bf II. THE NONLINEARIZATION SCHEME}
%\section{The   nonlinearization  scheme}
%\label{sec2}

Combining  field $q$  and its conjugate $r $ in 
 a matrix as  $U^{(0)} =q\sigma^{+}+ r\sigma^- $, where
$\sigma^\pm= {1 \over 2} (\sigma^1\pm i \sigma^2) $ and $\sigma^a , a =1,2, 3 $
are standard $ 2\times 2$-Pauli matrices, we can clearly  express
 our starting   linear dispersive equations (\ref{lineq}) as
  \ $
  iU^{(0)}_t=\sigma_3U^{(0)}_{xx},\ \
  U^{(0)}_t=U^{(0)}_{xxx},
 $
%\ll{lin}\ee
 etc. These linear equations  in turn can be expressed   
by  easy observation  
     as a   {\it linear
flatness} condition  $ U^{(l)}_t- V^{(l)}_{ix}=0 $,
  which is  defined    by ignoring the  nonlinear 
commutator term between   Lax operators.
%\ll{linUV}\ee,
This naturally    identifies  the     pairs of naive Lax operators 
$(U^{(l)}, V_i^{(l)}), \ i=2,3,\dots $ for the consecutive higher order linear  equations as
\be U^{(l)}=i \lambda\sigma_3
+iU^{(0)}, \ \ \ \ \mbox {and} \ \
%\be
   V^{(l)}_2=
%2i\lambda^2 \sigma_3+
 \sigma_3 U^{(0)}_x, \ \
V^{(l)}_3=
%-4i\lambda^3\sigma_3+
iU^{(0)}_{xx} ,  
 \ll{UV0}\ee
 etc.  with $\rm {tr} U^{(l)}=\rm {tr} V^{(l)}_i =0, \ i=2,3, \ldots $, since $\rm {tr}
 \sigma_3=
\rm {tr} U^{(0)}=\rm {tr}(\sigma_3 U^{(0)})=0$.  
The significance of   parameter $ \lambda$, which
enters here only as an additional constant and
 can be ignored   without any loss, will  become clear in
the process.

The Next important point is to detect the scaling dimension of the 
Lax operators as well as to  identify their crucial  building blocks (BB),
 existence of which seem to have been 
overlooked  in most of the work including the AKNS
construction.
%sesh15
\vskip .8cm
\noindent {\bf A. Scaling dimension and building blocks of the Lax pair}
% III. SCALING DIMENSION AND THE BUILDING BLOCKS OF THE LAX PAIR}
%\subsection{}

We  consider  length $L$ as the fundamental dimension and 
 from  the  linear evolution equation 
with $ N$-th order dispersion: $q_t=q_{\underbrace{xx \dots x}_N} $
 read out  the  dimension $[\cdot ] $ of time  as 
   $[T]= L^N $. The Lax
operators, generating infinitesimal space and time  translations,
 as    evident  from  the Lax equations 
  $ \Phi_x=U\Phi, \ \Phi_t=V\Phi$, naturally should have  the  scaling  dimensions (SD)
as  
 $[U]= L^{-1} , \ [V_N]=
[T]^{-1}=L^{-N}. $
Assuming that the  linear Lax pair $U^{(l)}, V^{(l)}_N $
   share  the same  scaling dimensions as well as the same building blocks
as their nonlinear counterpart,   we find from the
  explicit    structure  (\ref{UV0}),    the SD 
  of their constituent  elements  as    
 $[\lambda]=[U^{(0)}]= L^{-1}$ and  at the same time
   identify  the building blocks (BB) for  both the  Lax operators as the
set
 $\{\ \lambda, U^{(0)} \ \}$ and its x-derivatives: $ U^{(0)}_x ,  U^{(0)}_{xx} ,  $
etc. , which enter in their construction 
  linearly.  
 Our crucial  conjecture is  that 
  these  linear Lax pair 
can be     nonlinearized by   adding 
all possible nonlinear combinations  of    the same BB,  guided  by the
dimensional argument.
Note that, the condition $[U^{(0)}]= L^{-1} $  
fixes       the SD for the
field in this case as  $[ q]=[ r]= {L}^{-1}$.
 However, since  a linear equation as such  can not put any
restriction on the SD of its  fields, this input is not unique and
we show below that  a different fixation  of the  SD  for the field would lead to
another construction of the Lax pair.
% since  for the x-derivative     $[\frac \partial {\partial x}]=[L]^{-1} $.

%%Kundu-Eckhaus (KE) type  equation from  the LS equation
%%by a change of variable $q \to \psi=qe^{\int | q|^2 dx'} $\cite{Kun84}, etc.
%Similar transformation

 Recalling that a nonlinear  variable change can   generate  from a
 given integrable system   other  gauge equivalent
 equations
% , e.g. Kundu-Eckhaus  equation from the NLS equation,
%Chen-Lie-Liu  equation from the DNLS equation etc.
 \cite{Kun84},
   we  concentrate here only  on  fundamental integrable equations
like NLS, KdV DNLS etc.
  belonging  to the AKNS or the  KN family, while
 the gauge equivalent     systems
 can be obtained trough simple transformations.

\vskip .8cm

\noindent {\bf III. GENERATION OF NONLINEAR INTEGRABLE SYSTEMS}

%\section{Generation of nonlinear integrable systems }

  The dimension of the field $[q] $
 and its conjugate $[r] $ as well as  $[\lambda] $, which
  are not fixed by
 the starting   linear equations (\ref{lineq}),  should be  given
in our nonlinearization scheme
   as input.
 We show that for   $[q]= [r] =[\lambda]  =L^{-1} $, we get the AKNS
systems,
 while 
$[q]= [r] =[\lambda] =L^{-\frac 1 2} $ yields  the KN  hierarchy.

\vskip .4cm

%\subsection
\noindent{\bf A. AKNS integrable hierarchy  }

We consider first the case  $[q]=[r]=[\lambda]=L^{-1} , $ 
 which  gives  through  the above  construction: $
[ U^{(0)}]=[U^{(l)}] =L^{-1} $. We intend to 
  construct  now the space-Lax
 operator $U(\lambda) \in sl(2), \rm {tr} U(\lambda) =0 $, naturally with    $[U(\lambda)]=L^{-1} $,
     out of  $U^{(l)} $ by using 
  the BB: $\{ \lambda , U^{(0)} \} $.
Therefore  the only possibility left from the dimensional argument, which
only allows  summation of the terms having the same dimension,   is 
 to take   
 $U(\lambda)= U^{(l)}=i( \lambda\sigma_3
+U^{(0)}), $  recovering thus   
 the well known AKNS Lax operator \cite{solit1} uniquely.
 However for constructing  the
corresponding time-Lax operator  $ V_2(\la)\in sl(2)$ with $ \rm {tr}V_2(\lambda)=0 $
 and  SD: $L^{-2} ,  $
  in addition to  $V^{(l)}_2=
 \sigma_3 U^{(0)}_x, $
 nonlinear terms $ V^{(nl)}_2 $ with  the same   SD:
$L^{-2}$ are to be   constructed 
 out of    the same BB 
  $\{ \lambda , U^{(0)} \}$,  through  their nonlinear  products and powers like
 $\la  ^2, \la  U^{(0)}, (U^{(0)})^2  , $ following  the dimensional argument. Note that since $ [\partial_x ]=L^{-1}$,
 the derivative term $ U^{(0)}_x$  can appear  only in the linear
part, again from the   dimensional analysis. 
  
%=[\la ] ^2= [\la ] [U^{(0)}]=[(U^{(0)})^2] $.
  Therefore this set  of nonlinear terms 
is sufficient to construct  the nonlinear part of the time-Lax operator: 
$V^{(nl)}_2 =i(k_2\lambda^2
 \sigma_3+ k_1\lambda U^{(0)} +k_0\sigma_3 (U^{(0)})^{2}) $, upto
the integer coefficients $k_n, n=0,1,2 $. 
Note that fixing   the dimensionless integers $k_n $ 
with different terms in $V_2(\lambda) =V^{(l)}_2+
V_2^{(nl)} $ goes
 beyond the scope of
 the dimensional argument, which however
is achieved from the   flatness condition of  $
U(\lambda),V_2(\lambda)$,   yielding   several consistency
relations at different powers of  $ \lambda$. Interestingly, the
number of such relations obtained is just sufficient to determine the
numerical coefficients of all the terms 
  without any ambiguity. For example, it is easy to check that,
 the condition (\ref{flatness})
  for $U(\lambda),V_2(\lambda)$  yields three  
consistency
conditions at three different 
 powers of $\lambda^n,\ n=0, 1, 2  ,$  
of which  the condition at $n=0$ fixes $k_0=-1 $,  that at $n=1 $ fixes   
the integer $k_1=2 $, while  $n=2 $  gives the relation $k_2=k_1 $.
Thus all the integer coefficients are
 obtained  exactly,   
%together with the condition  ${\rm {trace}} V =0  $,  
  determining  
 the structure of the nonlinear part unambiguously as
 \be V^{(nl)}_2 =2i\lambda^2
 \sigma_3+ 2i\lambda U^{(0)} -i \sigma_3 (U^{(0)})^{2},
\ll{V2akns}\ee
 with $ \rm {tr}V^{(nl)}_2=0 , $ which constructs  the complete 
 time-Lax operator uniquely as $V_2(\lambda)=V^{(l)}_2+
V_2^{(nl)} $ . 
%, as given in (\ref{V2akns}).
 Note that the asymptotic
solutions of the Lax equation giving $e^{\pm i \lambda x} $ clarifies the
physical meaning of parameter  $\lambda $ as the {\it momentum} and  justifies  
the choice of its SD:   $L^{-1} $.
At the final step we obtain   the integrable
nonlinear    
 equation we are searching for, again from the flatness condition,
% as a coefficient for $\lambda ^0 $.
  where  a nonlinear term
$2\sigma_3 ( U^{(0)})^{3} $ appears now in addition to the initial
linear equation $ iU^{(0)}_t=\sigma_3U^{(0)}_{xx}$.  For $r=\pm q^* $ this nonlinear term
 reduces to $ \pm 2| q|^2q $, which
 together with the starting
LS equation   yields finally the integrable NLS equation, completing our
nonlinearization process.
%  second integrable equation in the AKNS  hierarchy, which
% where we recover  the
%   AKNS Lax pair

The  next higher  order  linear equation $q_t=q_{xxx} $
follows a  similar procedure in the nonlinearization process.
The  space-Lax operator being the same, one has to build only the
time-Lax operator $V_3(\lambda) $ by adding to the naive  operator
  $V^{(l)}_3=
iU^{(0)}_{xx}  $ a  nonlinear part $
V^{(nl)}_3$ with SD $ L^{-3}$, which is to be   constructed again
  from the same BB : $\{ \lambda , U^{(0)} \}$ and its derivative $ \ U^{(0)}_x, $ 
following our  conjecture.
Observe that the derivative term
can appear now in the nonlinear part on dimensional ground. By 
  all possible
nonlinear combinations as powers and products of the BB with total  SD = $ L^{-3}$ and
maintaining
$  \rm {tr} V_3(\lambda)=0$, 
 we can, similar to the
 above, construct uniquely  
%$3=[\la ][U^{(0)}_x]=[\la ][(U^{(0)})^2]=[\la ^2 ][U^{(0)}] =
%[U^{(0)} ][U^{(0)}_x]=[(U^{(0)})^3]$.
% This guiding  principle along with
%the condition $V^{(nl)}_3  \in sl(2) $ generates
 : 
%\bibitem{Vn23}
%AKNS: $ V_3(\lambda) = V_3^{(l)}+ V_3^{(nl)}, \
%\tilde V_2= 2i\lambda U^{(0)} -i \sigma_3 (U^{(0)})^{2},\\
\be 
V_3^{(nl)}= -4i\lambda^3\sigma_3+2 \sigma_3(
-U^{(0)}_x+i(U^{(0)})^2) \lambda   -4i  U^{(0)}\lambda^2+2i
(U^{(0)})^{3} -[ U^{(0)}, U^{(0)}_x],
 \ll{v3nl} \ee
%\cite{Vn23},
 yielding   finally 
 $ V_3(\lambda)= V^{(l)}_3+  V^{(nl)}_3 $. The flatness condition of
the pair
 $U(\lambda),V_3(\lambda) $
fixes again  the integer coefficients  in   different terms of (\ref{v3nl})
 and yields
the integrable  equation, by adding only one nonlinear term
%\be
$ \ (U^{(0)})^{2}U^{(0)}_{x}\ $
%\ll{nl3terms}\ee
   to  the
starting  linear dispersive equation $\ 
  U^{(0)}_t=U^{(0)}_{xxx} \
  $.
  For $ r=\kappa=const. , \ q=u$  the nonlinear term reduces  to
$\kappa 6 uu_x$, while for  $ r=q=v$ to $6 v^2 v_x $, generating thus the
well known  integrable 
 KdV  and   mKdV equations, 
from the  third order linear equation we  started with.

Similarly one can continue building the hierarchy of integrable equations,
  starting from the arbitrary   higher  order linear
  equation $q_t=q_{{xx \dots x}} $  and  nonlinearize it 
 following the  
procedure as above and using the  same BB and similar  dimensional argument, as we
have conjectured.
Since  the space-Lax operator $U(\lambda) $ remains the same,
 the task is
 to construct only the  time Lax operator $V_N(\lambda)  $ with the 
SD $ L^{-N} $ for arbitrary $ N$,
out of    the same BB: $ \{ \la  , U^{(0)}\} $ by
 all possible    combinations
  like \be (\la ) ^k
(\frac \partial {\partial x})^l(U^{(0)})^m \ll{Vn1}
\ee
 with the  dimensional constraint
$k+l+m=N $.
 It is easy to check that
 this partitioning of $ N$ determines  the total number of terms  that can
 appear in $ V_N(\lambda) $ as $L= 1 +\frac 1 2 N(N+1) $, which
interestingly tallis with    our  result obtained above, giving $L=4$  for  the NLS with
$N=2$  and   $L=7$   for the KdV and  mKdV with $N=3$.
% Interestingly  even without any construction, this partitioning in the  integer powers guided by SD,
% considering their  values between  $0 $ and $N$ with the obvious
%constraint  $l=0 $, when $m=0 $, easily gives us a number
Thus we can estimate the
exact number of terms that should appear in any higher order Lax operator,
 without even
calculating
their explicit form. This unique feature of our nonlinearization scheme,
obtained from the  dimensional argument and the identification of the BB, is absent
in all other available methods.    
\vskip .4cm

\noindent
%\subsection
{\bf B. Comparison with the AKNS scheme}

It is true that, though we have presented a simple and 
 original   scheme for  nonlinearizing linear equations to integrable
systems,
 based  on  the
   dimensional analysis and
using  the  building blocks of  the Lax  pair, 
 we could reproduce so far   only 
the known integrable equations,  which are 
obtainable  also  through the AKNS scheme \cite{solit1}.
In-spite of this fact 
and  before  presenting  our    new result,  we intend
to show  that  in comparison with   the AKNS
proposal \cite{solit1},
 the construction of    time-Lax operator 
$V_N(\la ) $,  vital in generating
  integrable equations,   is significantly
  simpler in the present scheme. 

Recall firstly, that   in the  AKNS method  one starts with a given
nonlinear field
equation  and not from its linearized version, as done here.

Secondly, for obtaining the explicit form of 
 $V_N(\la ) $ in the AKNS method, one  has to expand it  in the powers of spectral
parameter: $V_N(\la )=\sum_{n=1}^N=\lambda ^n V^{(n)} $, where $ V^{(n)} ,
n=1,2 , \ldots $ are $N $ number of $2 \times 2 $ unknown matrices with 
 $4N $ number of  unknown  functions, some of them being complex.
In our  construction    on the other hand,   the unknown 
coefficients are only   $N $ in number and moreover they are only  
real integers. This is because   we have  already  identified 
  the building
blocks of the Lax operators made up  from  parameter $\lambda $,
 a known   matrix $ U^{(0)}$ and its derivatives, and  the dimensional
argument, which is our key ingredient, guides us to collect them in the right 
combination.

Thirdly, in  the AKNS scheme the $4N $  
unknown    functions  are to be determined
 by   solving different  partial differential equations.
Our $N $  unknowns being only real integers are obtained  
from  simple algebraic equalities.

 This simplicity of our scheme becomes
more evident for higher values of $N $, where  $V_N $ contains 
$L=1+\frac 1 2 N(N+1) $ number of terms. Note that   determining a priori
 the number
of terms that should be present in any $V_N $, as we have shown above 
is again easy in our scheme but not in the AKNS.

Fourthly, while   for  
 the KN spectral problem the AKNS type direct expansion becomes
 even more complicated, our scheme 
covers  the KN system  in an unified way. Just a change in the scaling
dimension of the field switches our scheme from the AKNS to the KN family, 
as we   demonstrate
below.

Finally,    
as an important   application   we discover  new integrable hierarchies 
of  nonholonomic deformations for all members of   the AKNS family (see
Sec. V),
systematic account of which is absent in  AKNS \cite{solit1}.
\vskip .4cm

\noindent
%\subsection
{\bf C. KN integrable hierarchy }

 Our nonlinearization scheme , as we  show here,
is universally applicable 
for generating the KN hierarchy, which 
is not  readily available in the literature due to its apparent
complicacy.

We  follow again the same line of argument, making a slight change 
in the 
 input of the SD of the field as   $[q]=[r]=[\lambda]=L^{-\frac 1 2} $.
 This small deviation however    
  dramatically changes the outcome and   resolves   the  puzzle we faced in  generating uniquely
two different nonlinear equations, namely  the NLS and the   DNLS equations,
 starting from the same linear LS equation.
 
Note that   the SD of the linear Lax operator 
 has  been changed now to  $ [ U^{(l)}]= L^{-\frac 1 2}$,
while that for  the  genuine space Lax operator must always be $[\tilde
U(\lambda)]= L^{- 1 } $.
Therefore  the nonlinearization induced by the dimensional argument    
   should give $\tilde U(\lambda)=\lambda U^{(l)}=i \lambda^2\sigma_3
+i\lambda U^{(0)}$,
 which reproduces correctly   the well known  KN Lax
operator \cite{kn}. Note that, in principle, 
 one can also add on  dimensional ground. a term
like $i\sigma_3(U^{(0)})^2 $ to this Lax operator. 
 However, such an addition does not  produce
any independent  equation
that are not  derivable  from those obtained without its addition.

 For constructing the   corresponding time-Lax operator $\tilde V_N(\lambda) 
$ through  our nonlinearization,  one should remember 
 that the situation is a bit  different here, since  though the
 dimension of  $\tilde V_N(\lambda)$ must  remain   as   $ L^{- N } $,
the SD of its  BB: $ \{ \la , U^{(0)} \}$  has been changed.

Let us consider first the case with 
 $N=2 $ and start again from the  linear Schr\"odinger    equation. We repeat   
  the above  procedure for the  NLS  case, 
but    with the present change in the  dimension  of  the  BB, which should lead to 
the construction $\tilde V_2(\lambda)=\lambda V_2^{(l)}+ V^{(nl)}(\lambda) $
due to the dimensional constraint
   $  L^{- 2}$, where $ V_2^{(l)}= \sigma_3U^{(0)}_x $ is the naive linear
operator  as  in (\ref{UV0}).
 The nonlinear part $ V^{(nl)}( \lambda) $ should be   constructed  
 therefore by nonlinear combinations of  the BB with  dimension
 \be  L^{- 2}=
%[\la ][U^{(0)}_x]=
[\la ][(U^{(0)})^3]=[\la^2][(U^{(0)})^2]= [\la^3][U^{(0)}]=[\la ^4 ].\ee 
 Note again that though terms like
$i\sigma_3(U^{(0)})^4 , \ [U^{(0)},U^{(0)}_x] $ match in
dimensionality and hence can also be added, they
% give no independent equations and
  only yield    equations that  are derivable from those, obtained
without their addition. 
 This
identification
  together with   $\tilde V_2(\lambda) \in sl(2), \ \rm{tr} \tilde
V_2(\lambda)=0,$
constructs the
 time-Lax operator
%\bibitem{Vnn23} KN:
%204
\be
 \tilde V_2 (\lambda) = ( \sigma_3U^{(0)}_x+ i (U^{(0)})^{3})
\lambda- i \sigma_3(U^{(0)})^2 \lambda ^2+2iU^{(0)}\lambda
^3+2i\sigma_3\lambda ^4 ,  
 \ll{Vnn23}\ee
%\cite{Vnn23},
  uniquely , with  the integer coefficients 
%coefficients of individual terms
 fixed  as above from the consistency condition. This 
condition   generates also 
 the nonlinear equation with the integrable   nonlinearity  
  $i ((U^{(0)})^{3})_x $, which  for
$r=\pm q^* $   adds a term   $ \pm 2 (| q|^2q)_x$
  to our starting  LS equation $iq_t-q{xx}=0 ,$ yielding   the
    DNLS equation, as required.

Similarly the  higher order linear equations  we considered  above
nonlinearize now  to a different set of integrable equations
belonging to the KN hierarchy, due to the  changed  SD of the BB. Without
giving the details
%, which follow the
% same procedure  and the argument as above,
  we just mention,  that the construction,  though a bit tedious, is quite 
similar and straightforward and given again in the form (\ref{Vn1}),  with
 the total number of terms appearing in  
 $\tilde V_N(\lambda)$ as $\tilde L=1+N^2
$,  with arbitrary $N$ .
Note that this number of terms $\tilde L$, resulting from    
  the new  SD constraint:
 $\frac 1 2 (k+m)+l=N $,  is much higher 
  than the number $L$ appearing  in its  AKNS counterpart, showing why the KN systems are more
complicated than those of the AKNS.  
    Check that   $N=2 $ gives $ \tilde
L=5$ as we have  obtained above for the DNLS case.
%(\ref{Vnn23}).
%, for the on the integer
%powers,  due to the   changed SD input.

%\section{New integrable hierarchies for the AKNS family 
%with nonholonomic deformations}
%\label{sec4}

\vskip .8cm

\noindent {\bf IV. NEW INTEGRABLE HIERARCHIES FOR THE AKNS FAMILY WITH
NONHOLONOMIC DEFORMATIONS}

Note that the  emphasis  in the above nonlinearization scheme  is to construct
the time-Lax operator $ V_N(\lambda)= V^{(l)}_N+ V^{(nl)}_N (\lambda), $ by
building up the  nonlinear part involving positive powers of $\la $ and
confining to the elements only from the identified building block $ U^{(0)}
=q\sigma^++ r \sigma^-$ , 
which means to construct the  nonlinear combinations of the same basic
fields
$q,r $  and their
derivatives. Now
 we intend to   extend this construction
further by going beyond the conjecture of the  BB and 
introducing new perturbative functions with SD $L > N$. Thus keeping  $
U(\la)$   same as above,  we   add
 a nonlinear deformation $V^{(def)}_N (\la ) $ to the $V_N (\la ) $
by including   negative  powers of $\la ^{-n}, n=1,2,\ldots $
 in the dimensional argument, which we have ignored so far.
  This simple
extension, as we show, would
 discover a completely  new
class of integrable perturbed   equations with   hierarchy of 
higher order
nonholonomic deformations. Thus it creates a novel two-fold integrable 
hierarchy \cite{kunjpa082} by  adding a deformation hierarchy  to each of 
the known AKNS hierarchy.
 The
 well known equations like the  NLS, the  KdV, the mKdV and the SG  
 can be  deformed by perturbing functions
 with higher and higher order
nonholonomic constraints, preserving the original integrability.
The present result thus
shows a two-fold universality for the  nonholonomic deformations,
   found  recently 
for the KdV equation  \cite{6kdv,kup08,dkdv08}, since 
 one can   cover now the entire 
  AKNS  family , originating  from 
   this  single model, 
%, e.g.  NLS, KdV, mKdV, SG  etc.
 and at the same time can
  discover   a  new
  integrable deformation  hierarchy   
 for each members  of this family.

 It should be noted  in this context, that
the use of negative powers  of the spectral
 parameter in the time-evolution operator      
was  considered also  in some earlier occasions in 
the long history of integrable systems. However this  was either limited
\cite{solit1,maxBloch},  partial \cite{hmkdv} or camouflaged \cite{sourceI}.
The novelty of our result lies in the fact, that  within this extremely
 well
studied field we could discover in a  simple way
 a class of new integrable systems with 
a novel two-fold integrable hierarchy, which can
be interpreted as perturbed equations with integrable nonholonomic
constraints.
In this  construction of the time-Lax operator, we do not confine to the
BB,
which 
involve only   the  basic fields, but  include
a series of new perturbing functions, each
 influencing the basic field.
This situation can simulate  an interesting device for controlling the basic solitons
through multiple intervention, though  remaining within the 
scenario of the  exact solvability.
Such an idea was partially realized in the fiber optics communication
through doped media \cite{prl91}.

For presenting our new equations we start   with $V_2
(\la) $ for the AKNS system constructed above 
and    deform  it  first   by $V^{(def)}(\la )=\frac i 2 \lambda^{-1}G^{(1)}$,
%(a\sigma_3+G) $, where $a(t,x) $ and  matrix
% $G=g(t,x)\sigma_++ +\bar g(t,x)\sigma_- $
 with    a matrix function $ G^{(1)}$ with $\rm {tr}G^{(1)}=0 $ and   SD $  L^{- 3 } $.
 Since our intention is to introduce new perturbing function $g$
into the system, 
 we  derive from the flatness  condition
its structure  
 as an  integrable deformation of the
 second-order AKNS  equation:
  \be   iq_t-q_{xx}-2 (qr)q=g , \ll{nlsTs}\ee
    with the perturbing function, given by the matrix element 
$  g= G^{(1)}_{12}  $, subjected
 to  the  nonholonomic  differential constraint:
\be
% iU^{(0)}_t- \sigma_3 (U^{(0)}_{xx}+2 (U^{(0)})^{3})=\frac 1 2 [ \sigma_3,G]
 G^{(1)}_x=i[U^{(0)},G^{(1)}] .
 \ll{E} \ee
%where  the source  is the  matrix element  $g(t,x)= G_{12} $.
%, while $ G_{21} $ serves as the source for the conjugate equation.
%Notice that (\ref{nlsTs}) with (\ref{E}) can also be viewed as a new
%  source equation different from Melnikov's \cite{sourceI} and  
Remarkably, we can include further   perturbation into the system, which
would
interact with the basic field as well as with the initial perturbation. 
This process
can also be viewed   as  
% unlike  Melnikov's sources 
putting    higher order differential constraint on the original
perturbation.
To
achieve this higher  perturbation we 
 extend $V^{(def)}(\la )$ with  another deforming
 term $\frac i 2 \lambda ^{-2}G^{(2)}$, where $G^{(2)}$ is  a
 matrix function  of  SD
 $L^{-4} $.
%(b\sigma_3+F) $ with  $b(t,x) $ and  matrix  $F=f(t,x)\sigma_++
%+\bar f(t,x)\sigma_- $  and
Integrability condition (\ref{flatness}) now leads   to 
a further deformation of  (\ref{nlsTs})
given by the  higher order constraint    
%This   results to  the same NLS type equation with a
%source,  though the source is having  now a different   dynamical
%nature given by
 \be   G^{(1)}_x=i[U^{(0)},G^{(1)}]+i[\sigma_3, G^{(2)}],\ G^{(2)}_x=i[U^{(0)},G^{(2)}].
 \ll{EE} \ee
Generating such  higher   order integrable perturbations can be
 continued recursively by adding in $V^{(def)}(\la )$ more  and more terms
as
 $\frac i 2 \lambda ^{-j}G^{(j)}, j =1,2, \ldots n ,$ with arbitrary $n$.
New matrix functions $G^{(j)} $ have the scaling dimension
 $j+2 $ anf  $\rm {tr}G^{(j)}=0$. This  
 would 
result to  a new integrable hierarchy of nonholonomic   deformations 
for (\ref{nlsTs}), given recursively as
  \bea
 G^{(1)}_x&=&i[U^{(0)},G^{(1)}]+i[\sigma_3, G^{(2)}], \ \ldots \
 ,
\nonumber \\G^{(n-1)}_x&=&i[U^{(0)},G^{(n-1)}]+i[\sigma_3, G^{(n)}],
\nonumber \\ G^{(n)}_x&=&i[U^{(0)},G^{(n)}],
 \ll{EEn} \eea
 which clearly reduces to  (\ref{E}) for $n=1$ and to (\ref{EE}) for $
 n=2$.

%Note that, by considering the nonholonomic  deformation (\ref{nlsTs}),
%   as a source equation, system  (\ref{EEn})
%  can be represented as a novel  integrable   hierarchy of 
% source equations, the   possibility of which is surely absent     
%in  the traditional  scheme of Melnikov \cite{sourceI}.

Exactly in a similar way we can build up the hierarchy of integrable
perturbations for each member   in the known
    AKNS hierarchy of  higher nonlinear equations,
 which would lead thus to a novel two-fold integrable
hierarchy. In one the same nonlinear equation is perturbed by a function
with increasingly higher order differential constraints and in the other 
all different  higher  nonlinear equations are deformed by the same perturbing
function.

The most  important fact  about the hierarchies of all  perturbed
equations thus generated  is that, they are completely
  integrable systems and are exactly solvable by the inverse scattering
method (ISM), yielding $N $- soliton solutions.  This follows from the fact
that, for constructing such nonholonomic deformations, we have started from
the Lax pair, keeping the space-Lax operator $U(\la ) $ 
 same as the original one, while deforming the  time-Lax operator 
$V(\la ) $. Therefore the scattering problem, which is central to the ISM 
remains the same, while the time evolution of the spectral data 
only gets changed for the deformed models.
Consequently, in such perturbed systems
 along with the 
exact soliton solution for the basic field we can find  the 
exact solution for the perturbing function, which intriguingly 
takes also   the  solitonic form.
Moreover,
 one can analytically study the soliton dynamics as well as the scattering
of multiple solitons, a task impossible to carry out under usual
perturbation with a known function. 
It is  of  significant 
practical importance that, though  in such perturbed integrable systems,
the exact ISM is applicable  for extracting the soliton
solution,  we can
 bypass this involved and lengthy procedure and achieve the same result 
by  taking the well-known soliton solutions of the undeformed system
and  deforming them by suitably choosing their time-evolution, given by the
deformed soliton velocity \c{kunjpa082}.

 Remarkably,  a perturbed  soliton in the present
set up  behaves 
 like a particle
driven by   a  force and  exhibits
  an accelerated or decelerated motion,  depending on the nature of the 
deformation. Note that 
 unlike  known situations 
\cite{nisospec}   the variable  soliton velocity occurs here
 without any apparent inhomogeneity 
and  within the framework of an isospectral flow.
%in the usual way, since their spectral
%problem is not changed. However the time evolution of scattering
%data $b(\lambda,t), \ b_j(t) $ may change with additional terms like
%$e^{i{\frac {c^{(n)}} { \lambda ^n}}  t }$, depending on the
%asymptotic values $ c^{(n)}$ of the source term at space infinities.
%Consequently, The soliton velocity and the enveloping wave frequency
%may get boosted under the influence of the source (see Fig. 1).

\vskip .8cm

\noindent {\bf V. NEW INTEGRABLE EQUATIONS: SIMPLE EXAMPLES}
%\section {  New integrable  equations: simple examples  }

Using the above construction suitable for  equations belonging to the
AKNS spectral problem we can now analyze in  detail the new class
of integrable
perturbations for all members of this family, namely the  KdV, the mKdV, the NLS
and the  SG
equations. In particular  constructing the  
    matrix  Lax pair for each of them  we can   
 find  the  explicit form for all those  deformed equations, explore
their 
 higher deformations and most importantly,
 obtain their  exact N-soliton solutions.

Along with these perturbed equations we can  also study  their two-fold 
  integrable hierarchies, as  developed above.
 However we present here only the simplest equations in this  hierarchy,
which are  the most important
 cases involving the perturbation of the  well known  equations and
represent the lowest order  nonholonomic deformation, obtained from   
 (\ref{EEn}) with  $n=1$ or $2$.

\vskip .4cm

%\subsection 
\noindent{\bf A. Integrable perturbation of the   KdV equation}

Recent findings of the integrable
nonholonomic  deformation of the KdV equation,  equivalent to a 6th-order KdV
equation \cite{6kdv}, which  apparently contradicts  the 
accepted notion of nonexistence of any    even-order  equation in the KdV
hierarchy,
  arose considerable interest \cite{kup08,dkdv08}.
However, the exact integrability of this system through   
 AKNS type matrix Lax operator  or its exact N-soliton solutions 
through  the ISM could not be established. Consequently,
the integrable hierarchies allowed  by this system as well as the  
 dynamics of the solitons, including their nature of  scattering could not
be explored.

We on the other hand have  developed the  formalism  in Sec. IV for constructing 
  AKNS type Lax operators 
for all of its members  with nonholonomic deformations,  which should
 solve  completely the remaining unsolved aspects of the
 deformed KdV equation, as we show below.  

Recall that in the AKNS hierarchy  KdV type    
equations appear at      $N=3$ level, for which we have derived explicitly 
the time-Lax operator $V_3(\la ), $
 through our nonlinearization scheme. 
Now we intend to deform this Lax operator   by an additional
  $V^{(def)} (\la )$ matrix, which would result to a deformed  
 3rd-order AKNS equation 
 %$\lambda ^{-1}G_x +\lambda ^{-2} \tilde F $, which would result to an
\be q_t-q_{xxx}-6 (qr)q_x=g_x, \ll{3aknss}\ee
% U^{(0)}_t-   U^{(0)}_{xxx}-3
%((U^{(0)})^{2}U^{(0)}_{x}+U^{(0)}_{x}(U^{(0)})^2)=[\sigma_3, G]
with the multiple nonholonomic constraints as in (\ref{EEn}) on  
 the deforming
matrices  $G^{(j)} $,   having  higher scaling dimension: $j+3 $.
For deriving  the  perturbed KdV equation  
   we have to make the standard reduction 
    $r=1$ , $ q=u $ in  (\ref{3aknss}) to  yield
\be u_t-u_{xxx}-6
uu_x=g_x(t,x). \ll{kdvg}\ee
 For fixing now the  lowest order   constraint on the
deforming function, we have 
to add
  two  terms  with  $n=1,2 $  to
$V^{(def)}(\la)=\frac i 2 (\lambda ^{-1}G^{(1)}+\lambda ^{-2}G^{(2)}) $.
 It is important
to observe  
that adding only one deforming  term  in this case  gives trivial result,
while the double-deformation  yields  from  (\ref{EE}) the explicit
structure
\bea
G^{(1)}&=& (g+c) \sigma^3+i g_x \sigma^+\nonumber \\
G^{(2)}&=& -i \frac 1 2{ g_x}  \sigma^3+(g+c)   \sigma^-+i e\sigma^+
, \ e_x=iu g_x.
\ll{G12c} \eea
and the same compatibility condition leads to the nonholonomic 
constraint on the perturbing function as
%  the system 
\be  g_{xxx}+4u g_x+2u_x (g+c)=0. \ll{kdvEE}\ee 
Therefore the coupled system  (\ref{kdvg}-\ref{kdvEE})  
gives finally   the  integrable perturbation of the  KdV equation,
recovering  the recent result
 \cite{6kdv}.
We can find  now the exact  N-soliton solution for  this system of equations (\ref{kdvEE})  
through the application of the ISM. Referring to \cite{kunjpa082} for details we  
present here only   
the 1-{ soliton} solutions  for both the KdV field $u(x,t) $ and the perturbing 
function $g(x,t) $ 
as
\bea u(x,t)&=&\frac {v_0} 2 {\rm sech}^2 \xi,  \ \  \  \xi= \kappa(x+vt)+\phi, 
 \ll{1su} \\ 
 g(x,t)&=&c(t)(1- {\rm sech}^2
\xi), \ll{1sw} \eea 
  where $  v=v_0+v_d$, with $v_0$ being  the usual constant velocity of the KdV soliton, while 
$v_d=- \frac
{2\tilde c(t)}{v_0 t}$ with $\tilde c(t)=\int dt c(t) $ is its unusual
time-dependent part,
 induced by the deformation. The
 perturbing function,  itself taking  the solitonic form as (\ref{1sw}),
drives the field soliton  (\ref{1su})  through its 
asymptotic value $ \lim_{|x | \to \infty}g=c(t) .$ Therefore, in such
perturbed  integrable systems  one can control the motion of the  soliton from the
space-boundaries by
  tuning an  arbitrary
time-dependent function  $c(t) $, a feature seems to be
of immense physical significance    for
practical applications.
Fig. 1a shows the dynamics of this perturbed KdV soliton  moving with a constant
deceleration, 
 obtained for linear function  $c(t)=c_0 t $  with
$c_0<0$.
The exact  2-soliton solution  for  this perturbed  KdV equation, which
  can also be  derived explicitly,  shows beautiful elastic scattering 
of the accelerated solitons (see Fig. 1b).
\begin{figure}[!h]
\qquad \qquad
\includegraphics[width=5.cm,height=5. cm]{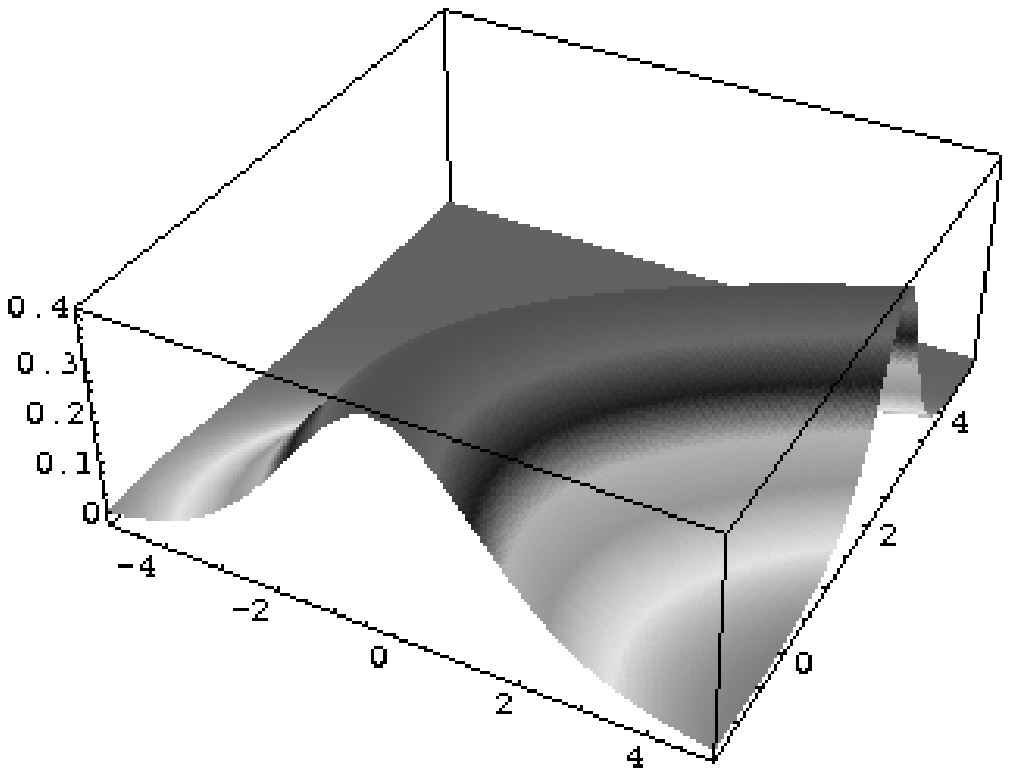}
 \qquad \qquad \ \ \ 
\includegraphics[width=5.cm,height=5. cm]{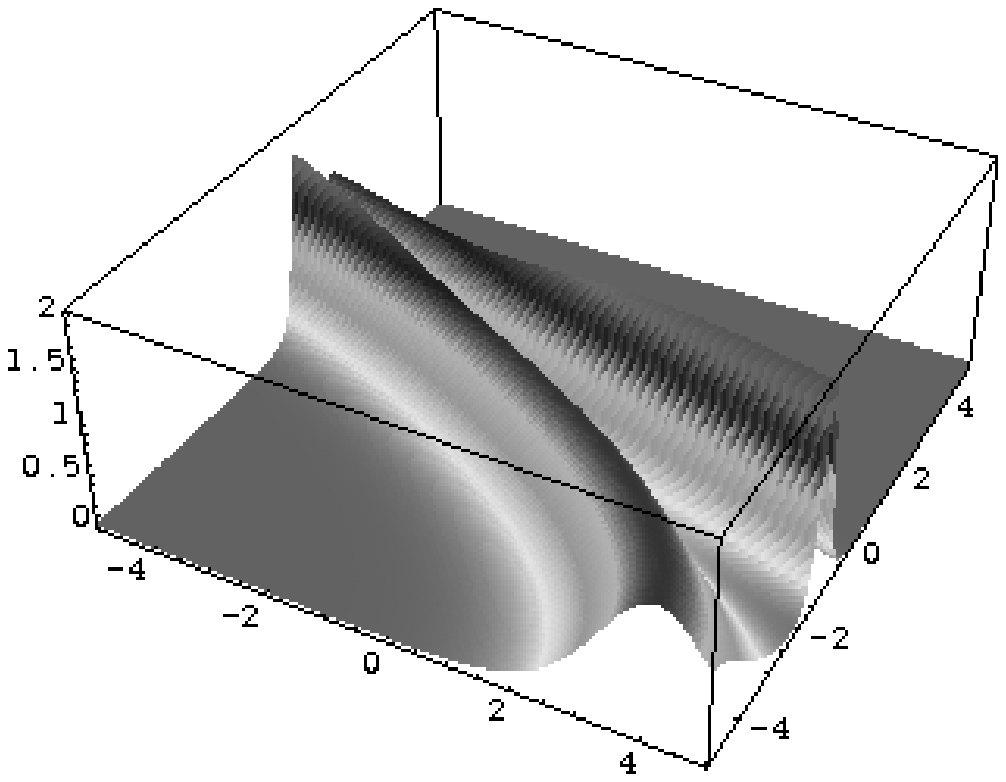}
%{s1skdvF.eps}
%\\ \hskip 2 cm  \qquad \qquad  (a) \qquad \qquad \ \ \ \ \ (b) \\
%\includegraphics[width=4.cm,height=3.1 cm]{sgd11s.eps}
% \put(-0.3, 11.9) {(b)}     \put(-14., 12.8) {(a)}
\caption{ Dynamics of the   exact soliton solution 
  of the    
integrable perturbation of the  KdV equation  for  the   field $ u(x,t)$. 
a) 1-soliton
  with the usual localized form but with 
an unusual deceleration, reflected 
in the bending of soliton trajectory in the (x,t)-plane, b)
  2-soliton 
 with the  usual elastic 
 scattering and phase shift, but  with  the 
 dynamics   dominated by an unusual  accelerated  motion.
}
%\end{center}
\end{figure}
%\subsection 

\noindent{\bf B. Integrable perturbation of the   mKdV
  equation}

Integrable equation with a nonholonomic deformation is   driven by
 a perturbing    function, which in turn is    subjected to
 a differential constraint.
 Therefore, analogous  to the deformed KdV, constructed above,
we can  derive a new 
 deformed  
mKdV equation
again from (\ref{3aknss}), but with  reduction $ q=r=v$.
Interestingly, in this case a single deformation with $n=1,  $ corresponding to
the constraint (\ref{E}) is sufficient to produce  the lowest perturbation we
are interested in, yielding the perturbed mKdV  
 \bea
& &v_t-v_{xxx}-6v^2v_x= w_x(t,x),\ll{dmkdva}\\
& & w_{xx}-2 v(c^2(t)-w_x^2)^{\frac 1 2}=0.\ll{dmkdvb}\eea
To see  the effect of  deformation 
        more closely
we construct   explicitly 1-{ soliton} solution, referring to \cite{kunjpa083} for the details of
the the details on the general  N-soliton.
The accelerating soliton of the perturbed mKdV, shown in Fig. 2, has the explicit form
\be v(x,t)=\frac {v_0} 2 {\rm sech}\xi,\ \xi= \kappa(x-vt)+\phi, \ v=v_0+v_d
\ll{1sv} \ee where $v_0=4\kappa^2 $ is the usual constant
velocity of the
 mKdV soliton, while
$v_d= \frac {2\tilde c(t)}{v_0 t}$ is the unusual time-dependent
part of the velocity, induced by the  deformation.
Note that  the perturbing function   $w(x,t) $   itself takes a self-consistent 
  solitonic form 

\be
w_x=c(t)\kappa {\rm sech}\xi {\rm tanh}\xi 
%\ \mbox{and} \ b=c(t)(1-\kappa ^2 {\rm sech}^2\xi  ).
\ll{wb1s}\ee
and  drives   the field soliton (\ref{1sv})
 to an  accelerated motion again through  
its asymptotic value $w(x,t)|_{|x | \to \infty}= c(t)$,
 sitting as a  forcing term at the space boundaries.

\begin{figure}[!h]

\qquad \qquad \ \qquad \includegraphics[width=5.cm,height=5. cm]{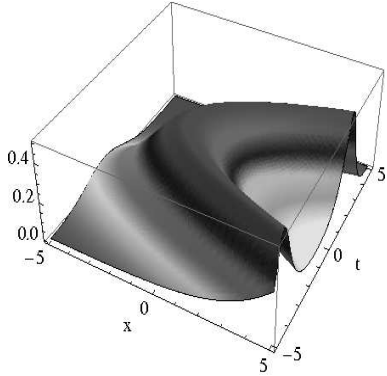}
%{mkdvmS.eps}
%{NHmkdvmS.eps}

\caption{  Exact  accelerated soliton  in the (x,t)-plane
  for the integrable perturbed mKdV equation.}
%(reflected in its  curvature with $c(t)=0.5 t $). }

\end{figure}
.
%  \end{figure}
%\subsection
\noindent {\bf C. Integrable perturbation  of the NLS equation} 

For constructing the integrable perturbation of the  NLS equation the stage is 
already set in (\ref{nlsTs}), since with a reduction   $r=q^* $ 
and   a minimum deformation     (\ref{E}), it
 would yield the NLS case
\be  iq_t-q_{xx}-2 |q|^2q=g , \
% iU^{(0)}_t- \sigma_3 (U^{(0)}_{xx}+2 (U^{(0)})^{3})=\frac 1 2 [ \sigma_3,G]
%\ G_x=i[U^{(0)},G]
\ll{NLSs1}\ee
with the perturbative function $g(x,t) $ subjected to a  nonholonomic
differential    constraint 
\be
g_x=-2ia q, \ a_x=i(q g^*-q^* g)
 \ll{Es1} \ee
where $G^{(1)}_{12}=g, \ G^{(1)}_{21}=g^*, \ G^{(1)}_{11}=-G^{(1)}_{22}=a $
are the deforming functions coupled through the basic fields $q, q^* $ . 
The set of constraints  (\ref {Es1})  may be simplified to a single
differential constraint
\be
\hat L(g)\equiv g_{xx}q-g_xq_x -2q^2(q g^*-q^* g)=0.
 \ll{Es11} \ee
 Eliminating  
the deforming function $g$ from (\ref{NLSs1}) and (\ref{Es11}) we can 
further derive
 a new 4th-order NLS equation expressed through the basic field as
\be
(q \partial_{xx}^2-q_x\partial_{x})(iq_t-q_{xx}-2 |q|^2q)
 +2q^2(i(|q|^2)_t+(qq^*_x-q^* q_x)_x)=0
 \ll{Es114}. \ee
As already mentioned, this perturbed system can  be solved exactly through
the
ISM. Remarkably however,  we can actually avoid the involved  procedure of
the ISM   in all such integrable nonholonomic
deformations and can  obtain   the explicit  soliton solutions for these
deformed equations by just suitably deforming the known undeformed
solutions. As a rule the perturbing
affect keeps  the form of the soliton intact, while it deforms some
parameters like velocity, frequency etc.,  making them in general time-dependent. 

More precisely, in  case of the  deformed NLS, 
 the original soliton  velocity $v_0 $  and the 
frequency $\omega
_0 $
of its enveloping wave are changed  with an addition of  
a  deformed  velocity  $ v_d(t)=\frac {\tilde  c(t) } {t |\lambda_1|^2}$  and the 
%$ \omega_0$
deformed  wave  frequency 
$ \omega_d(t)=-2\frac {\tilde c(t) \xi } {t|\lambda_1|^2}$,
where $\lambda_1 =\xi+i\eta$. The function
 $\tilde c(t) $ responsible for such deformation 
is linked  to  the asymptotic value of the perturbing function.
The accelerating NLS soliton qualitatively looks like that of the mKdV 
 when the dynamics of $|q | $ is plotted (see Fig.2).

We can generate again a two-fold integrable hierarchy for the perturbed NLS
system. The first one is
 the known hierarchy of higher order  NLS equations with
a perturbation  constrained
by the same nonholonomic deformation  (\ref{Es11}), while the second one 
is a new integrable hierarchy, where the same perturbed NLS equation 
(\ref{NLSs1}) is perturbed by  a hierarchy of 
   higher order deformations of the form (\ref{EEn}). 
 For example, the next  to the lowest order    deformation (\ref{Es11})
would
include another deforming function $e(x,t)$,
 in addition to  function $g(x,t)$
.
 Interestingly in this double-deformation 
the main  equation would remain same as  the  perturbed NLS equation  (\ref{NLSs1}),
while the constraint equation for $ g$ would be changed to
\bea
 & & \hat L(g) + 2i(eq_x-qe_x)=0, \ \nonumber\\
 & \mbox {or}\ & g_{xx}q-g_xq_x -2q^2(q g^*-q^* g) + 2i(eq_x-qe_x)=0,
\ll{Lge}\eea  coupling  to both the field $q$ and the  
 function $e $.
  The second deforming function $e$ in turn
is  constrained exactly  by the same differential constraint (\ref{Es11})
as  \be 
\hat L(e)=0, \ \  \mbox {or \ \ }
 e_{xx}q-e_xq_x -2q^2(q e^*-q^* e)=0. \ll{Le}\ee 
Even though  these new integrable perturbed NLS equations seem to be 
 rather academic, it is intriguing  that, a model like   our perturbed
system (\ref{NLSs1}-\ref{Es11}) is  implemented already in doped fiber for
efficient optical communication and therefore 
the implementation of  the  higher order integrable deformation discovered here 
 for a  a similar multi-doped  system 
seems to be  a promising possibility 
 \cite{KunPorz}.
\vskip .4cm

%SHESH1707

%\subsection
 \noindent {\bf D. Integrable perturbation   of the SG equation} 

Notice  that in the line of the present approach,  
the well known SG equation in the light-cone coordinates, which can be
obtained through reduction $ q=r=\theta_x$ from the AKNS system,  
 can itself be considered as  a
deformation  $ \theta_{xt
}=-i \tilde g$ of the linear wave equation   $ \theta_{xt }=0$,   
 with a constraint (\ref{E}) on the deforming function $\tilde g$. This constraint 
   can be     easily   resolved   
as
$
G^{(1)}_{12} =
\tilde g=i \sin 2\theta $, yielding      the standard SG equation: $ \theta_{xt
}= \sin 2\theta$ .
The situation  however becomes 
 more interesting if in the same  equation $ \theta_{xt }=-i\tilde g$, the deformation  $
\tilde g=ig$  
  is coupled to another   
 deformation $G^{(2)} $ ,
yielding 
the nonholonomic constraint (\ref{EE}) 
in the form
\be 
g_x=-2a \theta_{x}+2 e, \ a_x=2 \theta_x g, \ 
e_x=0
,\ll{Esg2} \ee
with the matrix elements  $G^{(1)}_{11}= a, \ G^{(1)}_{12}=-G^{(1)}_{21}= ig,\ G^{(2)}_{12}=G^{(2)}_{21}=e
(t),
 \ G^{(2)}_{11}= 0 $, where $ e(t)$ is an arbitrary function of $t$ .
 Expressing    (\ref{Esg2})
 through the perturbing function $ g$  
we obtain  the   nonholonomic deformation of the   SG equation as   
 \bea \theta_{xt} = g,  \ll{SGs0} \\
g_{xx}\theta_{x}-g_{x}\theta_{xx}
+4g\theta_{x}^3+2 e(t) \theta_{xx}=0
 \ll{sg2c} \eea
Note that solving the constraint (\ref {Esg2}) or equivalently 
(\ref {sg2c})
%  eliminating  $g $ from 
%(\ref{SGs0}) and  (\ref{sg2c})
 one  can get    different  
deformations of the  SG equation.  We derive one  such interesting   solution
 as 
\bea g&=&e(t)(\alpha \sin 2 \theta + \beta \cos 2 \theta ), \nonumber \\
  a&=&e(t)(\alpha \cos 2 \theta - \beta \sin 2 \theta). \ \alpha _x= \sin 2
\theta, \
\beta _x= \cos 2
\theta,  \ll{ga}\eea
which yields a new integrable deformation of the SG as
\be
\theta_{xt}= e(t)(\alpha \sin 2 \theta + \beta \cos 2 \theta ), \ll{sgE} \ee
 with $ \alpha, \beta$ as defined in (\ref{ga}). 
It is intriguing to note that, the new SG equation (\ref {sgE}) has an
additional $ \cos 2 \theta $ part along with the traditional $\sin 2 \theta $
term
and
the coefficient $e \alpha $,  which usually corresponds
 to the particle mass becomes a space-time dependent function.
It is worth mentioning that, though at $e=0 $ one can recover from (\ref{Esg2})  
the undeformed standard SG equation, as shown above, we 
 can no longer  obtain  the standard  SG equation from  
 its deformation (\ref{sgE}) at $ e \to 0$.
This new equation (\ref{sgE}), which is related to  the system considered
in \cite{hmkdv}, should however be examined carefully for the 
applicability  of the
ISM. Nevertheless, assuming the existence of its kink solution 
 we present 
in Fig. 3 the    accelerated kink
  for the  
field $\theta $ and the soliton solution  
for the  perturbing function for the integrable perturbed SG equation.
\begin{figure}[!h]

\includegraphics[width=5.cm,height=5. cm]{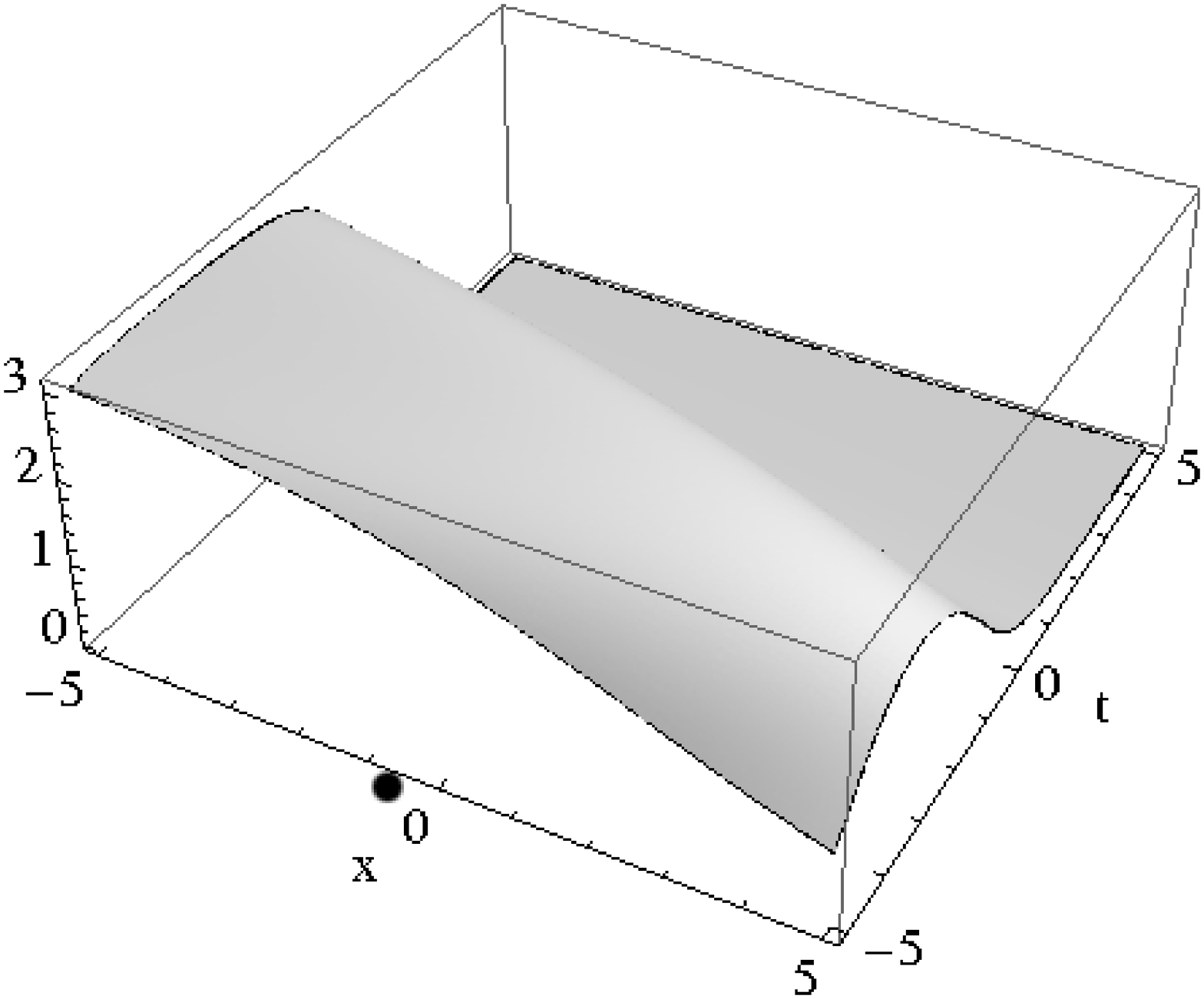}
%{NHsgths.eps}
\ \ \ \ \ \qquad \qquad
\includegraphics[width=5.cm,height=5. cm]{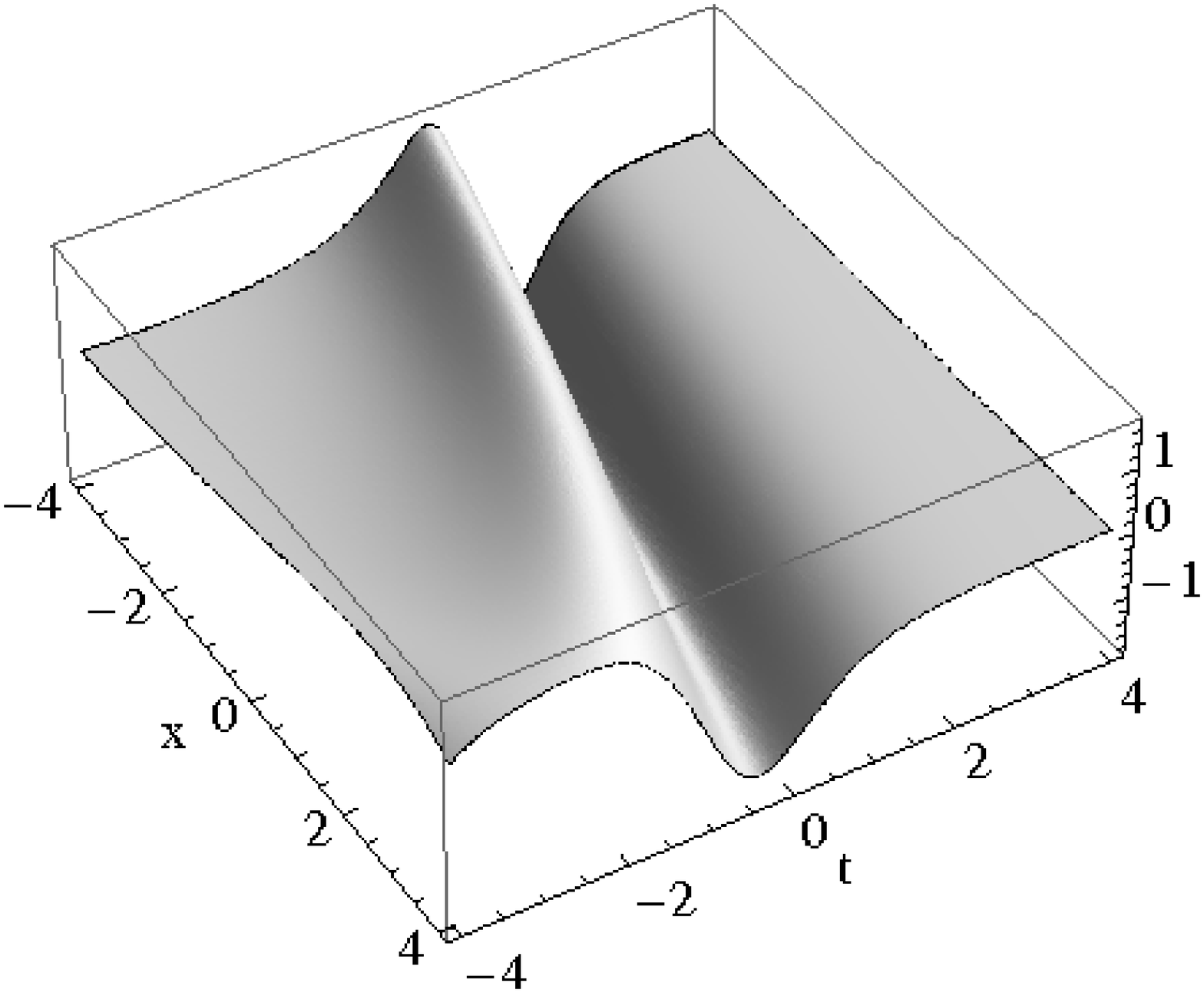}
%{NHsggs.eps}

\caption{  Exact soliton solutions for the 
integrable perturbation of the  sine-Gordon equation. a) 
Accelerating kink  solution  for   $ \theta(x,t)$. 
%with boundary force   $c(t)=0.5 t $
 b)  Perturbation  $g(x,t) $ in the solitonic form 
  with nontrivial time-dependent asymptotic   $c(t)=0.5 t $}. 
% a) The field and b)  the
%consistent perturbation  $g(x,t) $.
%\end{center}
\end{figure}

 Continuing the process of 
higher order differential    constraints,
 by introducing more and more 
terms with $\lambda ^{-n}, n=3,4, \ldots $  in $V^{(def)}(\la )$,   
  we can     generate  a    hierarchy of integrable 
  higher   nonholonomic 
 deformations in most of the cases.
In some cases, as the explicit calculations show, one has to add
 more than one higher terms to get the first nontrivial result.

%SHESH17072
%\section
\vskip 0.8cm

\noindent {\bf VI. PRESENT  NONLINEARIZATION SCHEME \& INTEGRABLE
PERTURBATIONS IN THE
BACKGROUND OF EARLIER RESULTS  } 
%{Present integrable perturbed systems   in the background of 
%earlier results}

We compared our  nonlinearization scheme for generating  
integrable
systems with the AKNS construction in Sec. III.B.
It is worth mentioning here  that the idea of this nonlinearization bears
some analogy as well as differences 
 with the well known Zakharav-Shabat (ZS) dressing method
\cite{ZS}. In the dressing method the Lax pair as well as 
the soliton  solution are constructed starting from the asymptotic
vacuum solution.
 In our nonlinearization scheme  on the other hand the Lax operators
are build up starting from the linear field equation, based only on the
dimensional argument and the notion of 
 the building blocks.
In analogy with the ZS dressing method   it is however
 tempting  to construct in our scheme  the soliton
solution for the nonlinear equation starting from 
 the solution of its  linearized equation.
% We unfortunately have not succeeded yet in achieving this goal.      
 
Extending our construction
further by nonholonomic deformation, we have gone  beyond the standard  AKNS
method and
considered perturbing matrices  of scaling dimension $L>N$  
in the $V_N(\lambda) $ operator. This amounts to extending this operator as
$V_{N,M}(\lambda)=V_N(\lambda) +V_{M}^{(def)}(\lambda) $ by
considering, together with the positive powers of $\la $ upto $N$, all its
 negative powers upto $M$.  
We  have shown through such deformations a beautiful
  universality of the recently found  nonholonomic
deformation of the KdV system \cite{6kdv,kup08}, by extending this concept to 
all members of the AKNS family, including the mKdV, the NLS and  the SG
equations. At the same
time we  have discovered a two-fold integrable hierarchy for  them,
with the explicit finding of some  interesting 
  multiple deformation.
Our construction allows the application of the exact 
ISM for obtaining  N-soliton solutions  for both the field and the 
perturbing functions,   which shows an
  unusual accelerated motion for the solitons.
%==========
However a significant 
 advantage of    such 
deformations is that, we can actually  avoid 
the application of the ISM in these cases and  construct
the  exact  soliton of the
deformed equations by  deforming the well known soliton 
solutions. Such perturbations only  deforms the time evolution of the
spectral data, which deforms in turn   
the soliton  velocity and for the complex fields also 
the  frequency of the enveloping wave.

%++++++++

It is true that in the long development of the theory of integrable systems,
the use of negative powers of $\lambda $ in the Lax operator $V(\lambda) $
was  implemented also in few  earlier occasions. However a consistent, systematic
and   full utilization  of this procedure
including the simultaneous presence of  higher positive and
 negative  powers of $ \la $,
associated to   all equations in  the AKNS family, 
and most importantly, the possibility of
introducing new perturbing functions as multiple deformations, which are our main emphasis
here, have never been undertaken. Similarly, the existence of a two-fold
integrable hierarchy  and  
 exact N-soliton solutions for both the field and the perturbing functions
in  all integrable perturbed systems of the AKNS family,
  as well as the possibility of   boosting the  soliton 
to accelerated motion  by the perturbing function
 through its  time-dependent  asymptotic 
 at the space boundaries, as revealed here, were  not explored or 
clarified in any earlier occasions.
%SHESH18222

As we understand, the AKNS themselves have considered integrable  SG 
and Maxwell-Bloch  systems going upto $\lambda ^{-1} $  \cite{solit1}.
A -ve hierarchy for the mKdV model was also considered  in
\cite{hmkdv}. However these work
either 
did not consider  both  +ve and -ve powers of $\la $ simultaneously, which
would result to the integrable perturbations, or they
went away from the perturbative models by considering only the equations for
the basic field. Thus these results did not come close to ours, though we
can derive them as particular cases of our more general integrable model.

 However an important physically applicable    model considered earlier
\cite{NLS-MB}  
  can be represented by the lowest order deformation of the NLS equation 
(\ref{NLSs1}-\ref{Es11}) found
through   our construction. This in turn indicates  the physical significance  
 of   our integrable  perturbed equations and opens up the possibility for 
application of the hierarchy of   perturbed    NLS equation, found here,   
to the nonlinear fiber optics communication through media with multiple
doping,
by switching to   higher order deformations \cite{KunPorz}. 

We should pay special attention to another class of interesting source 
equations proposed by Melnikov \cite{sourceI} and show that our  integrable 
perturbed models stand as complementary  to the Melnikov's models and can 
make contact with them    at a highly degenerate limit. In the Melnikov's formalism
  a set of  eigenvalues $\la _n, 1,2,\ldots, N$
 appear explicitly, which are    
  needed  to be distinct and strictly
nonvanishing  for any construction of   those models.
However as we show below, our construction could be related in fact  to the
complementary limit of the Melnikov's systems, when all these eigenvalues
become degenerate and moreover
 go to zero, simultaneously. Miraculously,   Melnikov's system,  in general too complicated,
  simplifies drastically  at this highly degenerate limit and reduces  to our simple perturbative model,
allowing exact  accelerating solitons  and suitable for physical applications.

Let us consider as a demonstration Melnikov's source equation for the NLS
system given as 

\be
iu_t+u_{xx}+2|u|^2u=\sum_n(\phi_n^2+\psi_n^2) \ll{MelNLS}\ee
and its complex conjugate, together with a  eigenvalue problem at discrete
eigenvalues $ \lambda_n, \ n=1,2,\ldots, N$ for a set of complex functions 
$(\phi_n,+\psi_n)$:
\bea \phi_{n,x}+u\psi_n &=& \lambda_n \phi_n
\nonumber \\
 \psi_{n,x}+u^*\phi_n &=&- \lambda_n \psi_n ,\ll{Melpsin}  
 \eea 
and their complex conjugates. As such this set  of $2 (1+2N)  $ number of 
 coupled complex 
equations is  evidently much more  complicated than our perturbed NLS 
(\ref{NLSs1}-\ref{Es11})
 and its   application to physical
models seems to be rather obscure. 
If however we  denote the {\it rhs} of (\ref{MelNLS}) by
$\sum_n(\phi_n^2+\psi_n^2)=g $ and the 
combination $\sum_n(\phi_n\psi_n-\phi_n^*\psi_n^*)=b $, then after some
algebraic manipulation  from  (\ref{Melpsin}) we find
\bea g_x+2ub&=&-2\sum_n(\lambda_n\phi_n^2-\lambda_n^*{\psi^*_n}^2),\nonumber\\
b_x+ug^*-u^*g&=& 0
\ll{gbMel}\eea
It is clear that the system  closes and   immensely simplifies only when
all eigenvalues become the same as well as  vanishing: $\lambda_n=0, n=1,2,\ldots,N $, which however is
opposite to the  Melnikov's theory, which strictly demands  all $\la _n \neq 0
$. We note immediately that at this vanishing limit from (\ref{gbMel})
 we get
\be g_x=-2ub
, \ b_x =u^*g-ug^*,
\ll{redMel}\ee
 which generates our
perturbed NLS (\ref{NLSs1}-\ref{Es11}) and for the real field reduction $u^*=u, \ g^*=-g $
reproduce further the  perturbed mKdV equation (\ref{dmkdvb}). Therefore we may conclude that 
at highly degenerate limit and in a situation  
 complimentary to the Melnikov's assumption, the well known source equations 
reduce to the lowest deformation cases of our perturbed equations. However the
possibility of generating higher order deformation for perturbing functions
constituting the novel integrable two-fold hierarchy of integrable perturbed
equations, we have discovered, seems to be   absent in the Melnikov's formalism.
%SHESH1912
%\section

\vskip .8cm 

\noindent{\bf VII. CONCLUDING REMARKS }
%Concluding remarks}

 The present scheme of nonlinearizing linear equations to integrable
systems, though bears similarity with the    AKNS   approach,
 differs in  its  motivation,   simplicity and   generality. Our procedure
  does  not start from a given integrable nonlinear equation,
 as customary  in the AKNS method, but
  aims to construct it by   nonlinearizing  a given  linear field
 equation 
 and      
is applicable  without much difficulty
to the AKNS as well as to the   KN family. Moreover due
 to identification  of the building blocks of the Lax operators and the 
effective use of the physical notion of 
 dimensional analysis, our construction   of the crucial 
 time-Lax operator 
 $V_N(\lambda)$
  becomes much simpler. In place    of 
  $4(N+1) $ unknown functions,  in general complex, in the AKNS scheme 
 obeying coupled partial  differential equations, 
 we have
 only $N+1 $ unknown real integers  to be determined from 
simple algebraic equalities.
 This
simplicity of our scheme becomes more evident at higher values of $ N$. 

 The  extension   of the notion of   integrability
from the KdV equation
% in late sixties
 to the  whole family of the AKNS system 
 was a major breakthrough achieved 
 in early seventies.
We now  find  a similar extension 
  for the  nonholonomic deformation of the KdV equation, discovered recently
\cite{6kdv,kup08}, to all 
 members of the AKNS family, which
 shows the universality  of such integrable deformations
beyond a   single known example \cite{6kdv}. 
Moreover, by discovering   an integrable  hierarchy 
 of increasingly higher order 
   deformation  for each of these cases, 
  we establish that this universality, in fact is two-fold.
One is 
 the known AKNS hierarchy
with   an additional fixed
deformation, while the  other
 is a new  integrable
hierarchy   for   each of the  NLS, the KdV, the mKdV and the SG  
 equations,  deformed by perturbing functions
 with increasingly higher order
nonholonomic constraints.
Interestingly, in the second  scenario  nonlinear integrable equations
can be generated by  perturbations only, even from a  linear dispersionless
 or trivial
 equation like $ q_t=q_x, \  q_t=0 $.

Such integrable perturbations are achieved  through an extension of our
 nonlinearization scheme, where we go beyond our
  building block conjecture and introduce new perturbative functions at
every higher step of deformation of the time-Lax operator,
 making  the fullest use of its  negative  power expansion
 in the spectral parameter. It seems that,
in spite  of numerous work in this well studied subject, these two simple
 steps were not considered earlier consistently, thereby missing a whole
class of integrable perturbed equations, which we discover here.
These integrable perturbed equations with  nonholonomic constraints 
 are  different and much simpler
than the   traditional source equations of 
 Melnikov. Interestingly however, they are related 
 in a exceedingly  subtle way, being complementary to each other and 
linked at a highly degenerate limit.
Nevertheless such a  relation can be shown to exist  only 
 at the lowest order deformation of our equations, while the novel hierarchy of
integrable deformation, that we find for our perturbed  systems, is totally
absent  in   Melnikov's source equations.

The  nonholonomic deformation inducing new integrable equations presented
here are  exactly solvable through the inverse scattering method.
%==========
However an  important  
 advantage of    such 
deformed integrable equations  is that, one can completely bypass the complicated 
ISM for extracting their exact soliton solutions and 
construct them by merely 
  deforming the well known 
solutions of the unperturbed equations. One finds that under  such perturbations
the form of the solitons remain the same, while 
the soliton  velocity and the wave   frequency  get additional terms,
capable of introducing  acceleration or deceleration in the solitonic motion.

%++++++++

The  explicit soliton solutions for the   simplest    integrable perturbed
   NLS, mKdV, KdV and  SG  equations,  
 presented here show  such unusual   accelerated
 motion forced by the perturbation.  Within this framework of  integrable
perturbation, one can also arrange these systems to
receive consistent multiple forcing, which  
are expected to have 
  significant physical applications, especially in  fiber optics
communication in doped media \cite{prl91,NLS-MB,KunPorz}.

Extentions of the nonholonomic deformation 
to the KN family of equations \cite{kun09}  as well as to $2+1 $ dimensional 
models are  exciting future problems. Another challenging problem, which we
could not solve here, is to 
construct the  soliton solution of the nonlinear equation
 from the  solution  of its   linearized
equation,
 by   using the present nonlinearization scheme, which might  
 define  some new
dressing method.

 \end{document}